\pdfoutput=1
\documentclass[conference]{IEEEtran}
\IEEEoverridecommandlockouts

\usepackage[utf8]{inputenc}

\usepackage{amsmath}
\usepackage{amssymb}
\usepackage{comment}
\usepackage{graphicx}
\usepackage{gensymb}
\usepackage{textcomp}
\usepackage{caption}
\usepackage{algorithm}
\usepackage{algpseudocode}
\usepackage{csvsimple}
\usepackage{array}
\usepackage{flushend}
\usepackage{balance}
\usepackage{multirow}
\usepackage{mathtools}
\usepackage{bm}
\usepackage{xcolor}

\usepackage[caption=false, font=footnotesize]{subfig}
\usepackage[bookmarks=false]{hyperref}
\hypersetup{
    colorlinks = true,
    citecolor  = blue,
    linkcolor  = blue,
    urlcolor   = blue,
}

\DeclarePairedDelimiter\floor{\lfloor}{\rfloor}
\DeclarePairedDelimiter\abs{\lvert}{\rvert}
\newcommand{\Expect}{{\rm I\kern-.3em E}}
\newcommand{\Enc}{\operatornamewithlimits{Enc}}

\newcommand{\HW}{\operatornamewithlimits{HW}}
\newcommand{\HD}{\operatornamewithlimits{HD}}

\newcommand{\LFSR}{\operatornamewithlimits{LFSR}}

\newcommand{\Gen}{\operatornamewithlimits{Gen}}
\newcommand{\Dec}{\operatornamewithlimits{Dec}}
\newcommand{\BER}{\operatornamewithlimits{BER}}

\providecommand{\innerprod}[1]{\langle#1\rangle}

\newtheorem{theorem}{Theorem}
\newtheorem{definition}{Definition}
\newtheorem{lemma}{Lemma}

\begin{document}

\author{\IEEEauthorblockN{Ye Wang, Xiaodan Xi, and Michael Orshansky}
\IEEEauthorblockA{\textit{Department of Electrical and Computer Engineering} \\
\textit{The University of Texas at Austin}\\
 Austin, TX, USA \\
\{lhywang, paul.xiaodan, orshansky\}@utexas.edu}
\thanks {© 2020 IEEE.  Personal use of this material is permitted.  Permission from IEEE must be obtained for all other uses, in any current or future media, including reprinting/republishing this material for advertising or promotional purposes, creating new collective works, for resale or redistribution to servers or lists, or reuse of any copyrighted component of this work in other works.}
}

\title{Lattice PUF: A Strong Physical Unclonable Function Provably Secure against Machine Learning Attacks}

\maketitle

\begin{abstract}
We propose a strong physical unclonable function (PUF) provably secure against machine learning (ML) attacks with both classical and quantum computers. 
Its security is derived from cryptographic hardness of learning decryption functions of public-key cryptosystems.  
Our design compactly realizes the decryption function of the learning-with-errors (LWE) cryptosystem. 
Due to the fundamental connection of LWE to lattice problems, we call the construction the lattice PUF.

Lattice PUF is constructed using a physically obfuscated key (POK), an LWE decryption function block, and a linear-feedback shift register (LFSR) as a pseudo-random number generator. 
The POK provides the secret key of the LWE decryption function; its stability is ensured by a fuzzy extractor (FE). 
To reduce the challenge size, we exploit distributional relaxations of space-efficient LWEs. 
That allows only a small challenge-seed to be transmitted with the full-length challenge generated by the LFSR, resulting in a $100$X reduction of communication cost.
To prevent an active challenge-manipulation attack, a self-incrementing counter is embedded into the challenge seed. 

We prototyped the lattice PUF with $2^{136}$ challenge-response pairs (CRPs) on a Spartan 6 FPGA, which required $45$ slices for the PUF logic proper and $233$ slices for the FE. 
Simulation-based evaluation shows the mean (std) of uniformity to be $49.98\%$ ($1.58\%$), of uniqueness to be $50.00\%$ ($1.58\%$), and of reliability to be $1.26\%$ ($2.88\%$). 
The LWE concrete hardness estimator guarantees that a successful ML attack of the lattice PUF will require the infeasible $2^{128}$ CPU operations. 
Several classes of empirical ML attacks, including support vector machine, logistic regression, and deep neural networks, are used: in all attacks, the prediction error remains above $49.76\%$ after 1 million training CRPs. 

\end{abstract}

\section{Introduction}
\label{sec:intro}
Silicon physical unclonable functions (PUFs) are security primitives widely used in device identification, authentication, and cryptographic key generation \cite{suh2007physical}.
Given an input challenge, a PUF exploits the randomness inherent in CMOS technology to generate an output response. 
In contrast to weak PUFs, also called physically obfuscated keys (POKs) using the taxonomy of \cite{herder2017trapdoor}, which supply limited amount of challenge-response pairs (CRPs), strong PUFs have an exponentially large CRP space.

\emph{In this paper, we propose a strong PUF that is secure against machine learning (ML) attacks with both classical and quantum computers.} 
As a formal framework to define ML resistance, we adopt the probably approximately correct (PAC) theory of learning \cite{mohri2012foundations}. 
Specifically, the PAC non-learnability of a decryption function implies that with a polynomial number of samples, with high probability, \emph{it is not possible to learn a function accurately by any means.} 
The main insight, which allows us to build such a novel strong PUF, is our reliance on the earlier proof that PAC-learning a decryption function of a semantically secure public-key cryptosystem entails breaking that cryptosystem \cite{kearns1994cryptographic, kharitonov1993cryptographic, klivans2006cryptographic}.
We develop a PUF for which the task of modeling is equivalent to PAC-learning the decryption function of a learning-with-errors (LWE) public-key cryptosystem.
The security of LWE cryptosystems is based on the hardness of LWE problem that ultimately is reduced to the hardness of several problems on lattices \cite{regev2009lattices}. 
The input-output mapping between the PUF and the underlying LWE cryptosystem can be briefly summarized as follows: challenge $\Longleftrightarrow$ ciphertext and response $\Longleftrightarrow$ decrypted plaintext. 
Notably, LWE is believed to be secure against both classical and quantum computers.
Because of the intrinsic relation between the proposed PUF and the security of lattice cryptography we call our construction the \textbf{lattice PUF}. 

The lattice PUF is constructed using a POK, an LWE decryption function block, a linear-feedback shift register (LFSR), a self-incrementing counter, and a control block.
The entire implementation is lightweight and fully digital. 

The LWE decryption function block is the core module of the lattice PUF, generating response (plaintext) to each submitted challenge (ciphertext).
Design parameters of the LWE decryption function in the lattice PUF are chosen by balancing the implementation costs, statistical performance, and the concrete hardness of ML resistance.
We develop a measure of ML security in terms of the total number of operations needed to learn a model of the PUF.
Such concrete hardness is established by the analysis of state-of-the-art attacks on the LWE cryptosystem  \cite{lindner2011better, micciancio2009lattice} and evaluated by the estimator developed by Albrecht et al. \cite{albrecht2015concrete}.
Using this estimator, we say that a PUF has $k$-bit ML resistance if a successful ML attack requires $2^k$ operations. 
We implement the LWE decryption function with guaranteeing $128$-bit ML resistance.
However, directly using a LWE decryption function as a strong PUF is not efficient since $1$-bit of response requires $1288$-bit input challenges.

We further develop an improved design for resource-constrained environments that dramatically (by about 100X) reduces the communication cost associated with PUF response generation.
This is achieved by \emph{by exploiting distributional relaxations allowed by recent work in space-efficient LWEs} \cite{galbraith2013space}.
This allows introducing a low-cost pseudo-random number generator (PRNG) based on an LFSR and transmitting only a small seed. 
Finally, while the focus of the paper is a PUF that is secure against passive attacks, we address the risk of an active attack by adopting the technique in \cite{yu2016lockdown}: we introduce a self-incrementing counter and embed the counter value into a challenge seed. 
This makes the attack impossible as the counter restricts the attacker's ability to completely control input challenges to the LWE decryption function.

We construct the lattice PUF to achieve a CRP space of size $2^{136}$.
Statistical simulation shows excellent uniformity, uniqueness, and reliability of the proposed lattice PUF.
The mean (standard deviation) of uniformity is $49.98\%$ ($1.58\%$), and of inter-class HD is $50.00\%$ ($1.58\%$).
The mean BER (intra-class Hamming distance (HD)) is $1.26\%$.
We also validate the empirical ML resistance of the constructed lattice PUF via support vector machines (SVM), logistic regression (LR), and neural networks (NN). 
Even with a deep neural network (DNN), which is considered to be one of the most powerful and successful ML attacks today, the prediction error stays above $48.81\%$ with 1 million training samples. 
The proposed lattice PUF requires a $1280$-bit secret key. 
A concatenated-code-based fuzzy extractor (FE) is utilized to reconstruct stable POK bits.
Assuming an average bit error rate (BER) of $5\%$ for raw SRAM cells, the total number of raw SRAM bits needed is $6.5$K, in order to achieve a key reconstruction failure rate of $10^{-6}$.  
We implement the entire PUF system (except for raw SRAM cells) on a Spartan 6 FPGA.
The PUF logic, including an LWE decryption function, a $256$-tap LFSR, a $128$-bit self-incrementing counter, requires only $45$ slices. 
The concatenation-code-based FE takes $233$ slices.
Compared to several known strong PUFs, the proposed PUF is significantly more resource-efficient.

\section{Background Work}
\label{sec:bg}
In order for a strong PUF to be an effective security primitive, the associated CRPs need to be unpredictable.
In other words, strong PUFs are required to be resilient to modeling attacks via ML. 
The question of whether it is possible to engineer a ML secure and lightweight strong PUF has been a long-lasting challenge \cite{vijayakumar2016machine}. 

SVM is utilized to successfully attack a $64$-bit arbiter PUF (APUF) in \cite{lim2005extracting}.
Subsequent modification of the original APUF aimed to strengthen ML resistance, including bistable ring PUF \cite{chen2011bistable}, feed-forward APUF \cite{lim2005extracting}, and lightweight secure PUF (LSPUF) \cite{majzoobi2008lightweight}, have also been broken via improved ML attacks \cite{becker2015gap, ruhrmair2010modeling, schuster2014evaluation, ganji2016strong}.
Recent proposed interpose PUF (IPUF) \cite{nguyen2019interpose} claims provable ML resistance by assuming XOR APUFs are hard to learn and rigorously reducing IPUF modeling to XOR APUFs.
Unfortunately, both their assumption and claims are proved wrong: \cite{DBLP:journals/iacr/SantikellurBC19} demonstrates that XOR APUFs and IPUFs are actually vulnerable to deep-learning-based modeling attacks.
There are often complex reasons why claims and rigorous proofs of security fail in practice. 
The most fundamental one is that their claims rely on recent conjectures made from empirical findings. 
In contrast, the security proof of lattice PUF is based on the hardness of several basic lattice problems, which are seen as foundational results in math and computer science, and are widely believed true. 

By exploiting higher intrinsic nonlinearity, some strong PUFs \cite{kumar2014design, xi2017strong} exhibit empirically-demonstrated resistance to a list of ML algorithms. 
Empirical demonstrations of ML resistance are not fully satisfactory since they can never rule out the possibility of other more effective ML algorithms. 

The so-called controlled PUF setting \cite{gassend2008controlled} attempts to ensure the ML resistance via cryptographic primitives such as hash functions. 
However, the use of hash functions inside a PUF endangers the promise of a strong PUF as a lightweight structure. 
Strong PUF constructions using established cryptographic ciphers, such as AES \cite{bhargava2014efficient}, have similar challenges. 

Recent work \cite{fuller2013computational,herder2017trapdoor,jin2017fpga} have also utilized lattice-based problems, including learning-parity-with-noise (LPN) and LWE, to realize computationally-secure FEs and, as a byproduct, to construct strong PUFs.
\footnote{A computational FE guarantees absence of information leakage from publicly shared helper data via computational hardness in contrast to conventional FEs that need to limit their information-theoretic entropy leakage.} 
The fundamental security property that \cite{fuller2013computational,herder2017trapdoor,jin2017fpga} rely upon is the \emph{computational hardness of recovering a private key from a public key in a public-key cryptosystem}. 
Their CRP generation is based on generating multiple private keys (playing the role of PUF responses) and multiple public keys (playing the role of PUF challenges). 
This is only possible because multiple public keys are derived using a fixed (same) source of secret POK bits, embedded in the error term of LPN or LWE. 
As was shown in \cite{apon2017efficient}, the fact that multiple CRPs have shared error terms can be easily exploited allows a computationally-inexpensive algorithm for solving an LPN or LWE instance, thus compromising the hardness of LPN or LWE problems. 
Thus, by itself \cite{fuller2013computational,herder2017trapdoor,jin2017fpga}, \emph{the resulting PUF does not have resistance against ML modeling attacks}. 
This vulnerability is fixed in \cite{herder2017trapdoor,jin2017fpga} by introducing a cryptographic hash function to hide the original CRPs, which violate the principle of lightweightness.
\emph{In stark contrast, the proposed lattice PUF derives its security by directly exploiting a distinctly different property of public-key cryptosystems: the theoretically-proven guarantee that their decryption functions are not PAC-learnable}. 
In the lattice PUF, the above-discussed vulnerability is absent since the publicly known challenges are ciphertexts and the security of the cryptosystem ensures that a fixed private key (the POK, in our case) cannot be recovered from ciphertexts.

\section{LWE Decryption Functions Are Hard to Learn}
\label{sec:lwe}
This section formally defines ML resistance of strong PUFs via the notion of PAC learning and shows why LWE decryption functions are attractive for constructing post-quantum ML-resistant PUFs. 
In this section, we focus on passive attacks in which the attacker can observe the challenges sent to the verifier but is unable to generate challenges of his or her choice.

\subsection{ML Resistance as Hardness of PAC Learning}
A strong PUF can be modeled as a function $f:\mathcal{C}\rightarrow \mathcal{R}$ mapping from the challenge space $\mathcal{C}$ (usually $\{0,1\}^n$) to the response space $\mathcal{R}$ (usually $\{0,1\}$).
We call $f$ the true model of a strong PUF since it captures the exact challenge-response behavior. 

ML attacks are usually performed by relying on a functional class of candidate models, collecting CRPs as the training data, and running a learning algorithm to obtain a model from the candidate class which best approximates the true model. In addition to the approximation quality, the criteria of evaluating the effectiveness and efficiency of the learning algorithm also include the sample and time complexity. 
To claim that a strong PUF is easy to learn, one can propose a learning algorithm which finds a CRP model with good approximation quality using a small number of sample CRPs and terminates in a short time.
The converse is difficult: 
to claim that a PUF is hard to learn, one must show that all possible learning algorithms fail to provide models with good approximation quality, or they require a large number of CRPs or a long running time.

We argue that the only known framework for seeking a provable notion of ML resistance with a formal analysis of approximation quality, sample size, and time complexity is the PAC learning model \cite{mohri2012foundations}.
We now formalize the passive modeling attack scenario in the context of PAC learning.
A PAC-term for a true model $f$ of a strong PUF is a concept.
Denote as $\mathcal{F}$ the set of all possible PUF-realized functions (every instance of a PUF creates its unique functional mapping $f$).
The set of candidate models used in the learning algorithm is the hypothesis set $\mathcal{H}$. The goal of a learning algorithm is to select a candidate model that matches the true model well. 
Importantly, as shown later, the proof of PAC-hardness guarantees that $\mathcal{H}$ does not have to be restricted to be the same as $\mathcal{F}$ of true models.
This generalization permits a stronger \emph{representation-independent} PAC-hardness proof. While not always possible, representation-independent hardness can be proven for PAC-learning of decryption functions ensuring that no matter how powerful and expressive the chosen $\mathcal{H}$ is, PAC learning decryption function requires exponential time. 

Within the PAC model, CRPs in a training set are assumed to be independent and identically distributed (i.i.d.) under a certain distribution $\mathcal{D}$.

We say a set $\mathcal{F}$ of strong PUFs is PAC-learnable using $\mathcal{H}$, if there exists a polynomial-time algorithm $\mathcal{A}$ such that $\forall \epsilon > 0$, $\forall \delta >0$, for any fixed CRP distribution $\mathcal{D}$, and $\forall f\in\mathcal{F}$, given a training set of size $m$, $\mathcal{A}$ produces a candidate model $h\in\mathcal{H}$ with probability of, at least, $1-\delta$ such that
\begin{equation*}
\Pr_{(\mathbf{c},r)\sim\mathcal{D}}[f(\mathbf{c})\neq h(\mathbf{c})] < \epsilon.
\end{equation*}

In conclusion, our strategy is to say that a strong PUF is ML-resistant if it is not PAC-learnable (i.e., that it is PAC-hard). 
PAC-hardness implies that any successful ML attack requires at least an exponential running time. 

\subsection{Decryption Functions Are not PAC Learnable}
What is critically important is that there exist functions that are known to be not PAC-learnable. 
Specifically, a class of decryption functions of secure public-key cryptosystems is not PAC-learnable, as established by \cite{kearns1994cryptographic,klivans2006cryptographic}. 
We outline their proof below.

A public-key cryptosystem is a triple of probabilistic polynomial-time algorithms $(\Gen,\Enc,\Dec)$ such that:
(1) $\Gen$ takes $n$ as a security parameter and outputs a pair of keys $(pk,sk)$, the public and private keys respectively;
(2) $\Enc$ takes as input the public key $pk$ and encrypts a message (plaintext) $r$ to return a ciphertext $\mathbf{c} = \Enc(pk,r)$;
(3) $\Dec$ takes as input the private key $sk$ and a ciphertext $\mathbf{c}$ to decrypt a message $r = \Dec(sk,\mathbf{c})$.
We only need to discuss public-key cryptosystems encrypting $1$-bit messages ($0$ and $1$).

One of the security requirements of a public-key cryptosystem is that it is computationally infeasible for an adversary, knowing the public key $pk$ and a ciphertext $\mathbf{c}$, to recover the original message, $r$. 
This requirement can also be interpreted as the need for indistinguishability under the chosen plaintext attack (also often referred to as semantic security requirement) \cite{katz2014introduction}. 
Given the encryption function $\Enc$ and the public key $pk$, the goal of an attacker is to devise a \emph{distinguisher} $\mathcal{A}$ to distinguish between encryption $\Enc(pk,r)$ of $r=0$ and $r=1$ with non-negligible probability:
\begin{equation*}
\abs{\Pr[\mathcal{A}(pk,\Enc(pk,0))=1] - \Pr[\mathcal{A}(pk,\Enc(pk,1))=1]}\geq \epsilon.
\end{equation*}
A cryptosystem is semantically secure if no polynomial-time attacker can correctly predict the message bit with non-negligible probability.

The connection between the above-stated security of a public-key cryptosystem and the hardness of learning a concept class associated with its decryption function was established in \cite{kearns1994cryptographic,klivans2006cryptographic}. 
The insight of \cite{kearns1994cryptographic,klivans2006cryptographic} is that PAC-learning is a natural result of the ease of encrypting messages with a  public key. 
Since the encryption function $\Enc$ and the public-key $pk$ is known, the distinguishing algorithm can sample independent training examples in the following way: (1) picking a plaintext bit $r$ uniformly randomly from $\{0,1\}$, (2) encrypting $r$ to get the ciphertext $\mathbf{c}=\Enc(pk,r)$. (We later refer to the resulting distribution of ciphertext as the "ciphertext distribution".)
Next, the distinguishing algorithm passes the set of training examples ($(\mathbf{c},r)$'s) into an algorithm for learning the decryption function $\Dec(sk,\cdot)$.
The PAC learning algorithm returns a model $h(\cdot)$ that aims to approximate $\Dec(sk,\cdot)$. 
Using $h(\cdot)$, one could distinguish between ciphertexts stemming from $r=0$ and $r=1$ with non-negligible probability. 
This would entail violating the semantic security of the cryptosystem. 
Technically, this can be summarized as follows \cite{kearns1994cryptographic,klivans2006cryptographic}.
\begin{theorem}
\label{thm:crypto_hardness}
If a public-key cryptosystem is secure against chosen plaintext attacks, then its decryption functions are not PAC-learnable (under the ciphertext input distribution).
\end{theorem}

\subsection{LWE Is Post-Quantum Secure}
\label{sec:lwe_crypto}
According to the cryptographic hardness above, decryption functions of any secure public-key cryptosystem, such as Rivest–Shamir–Adleman (RSA) and elliptic-curve cryptography (ECC), can be used to construct ML-resistant PUFs. 
However, integer-factoring-based cryptosystems, including RSA and ECC above, become insecure with the development of quantum computers. 
Among all post-quantum schemes \cite{bernstein2009introduction}, the LWE cryptosystem based on hard lattice problems appears to be most promising due to its implementation efficiency and stubborn intractability since 1980s. 

A lattice $\mathcal{L}(\mathbf{V})$ in $n$ dimensions is the set of all integral linear combinations of a given basis $\mathbf{V}=\{\mathbf{v}_1,\mathbf{v}_2,\ldots, \mathbf{v}_n\}$ with $\mathbf{v}_i \in \mathbb{R}^n$:
\begin{equation*}
\mathcal{L}(\mathbf{V}) = \{a_1\mathbf{v}_1 + a_2\mathbf{v}_2+\ldots a_n\mathbf{v}_n: \: \forall a_i \in \mathbb{Z}\}.
\end{equation*}

The LWE problem is defined on the integer lattice $\mathcal{L}(\mathbf{V}) = \{(\mathbf{a},\innerprod{\mathbf{a},\mathbf{s}})\}$ with a basis $\mathbf{V}=(\mathbf{I};\mathbf{s})$, in which $\mathbf{I}$ is an $n$-dimensional identity matrix and $\mathbf{s}$ is a fixed row vector (also called the secret) in $\mathbb{Z}_q^n$.
Throughout this paper, vectors and matrices are denoted with bold symbols with dimension on superscript, which can be dropped for convenience in case of no confusion.
Unless otherwise specified, all arithmetic operations in the following discussion including additions and multiplications are performed in $\mathbb{Z}_q$, i.e. by modulo $q$.

For the lattice $\mathcal{L}(\mathbf{V})= \{(\mathbf{a},\innerprod{\mathbf{a},\mathbf{s}})\}$ with dimension $n$, integer modulus $q$ and a discrete Gaussian distribution $\bar{\Psi}_\alpha$ for noise, the LWE problem is defined as follows.
The secret vector $\mathbf{s}$ is fixed by choosing its coordinates uniformly randomly from $\mathbb{Z}_q$. 
Next $\mathbf{a}_i$'s are generated uniformly from $\mathbb{Z}_q^n$.
Together with the error terms $e_i$, we can compute $b_i = \innerprod{\mathbf{a},\mathbf{s}} + e_i$. 
Distribution of $(\mathbf{a}_i,b_i)$'s over $\mathbb{Z}_q^n\times\mathbb{Z}_q$ is called the LWE distribution $A_{\mathbf{s},\bar{\Psi}_\alpha}$.
The most important property of $A_{\mathbf{s},\bar{\Psi}_\alpha}$ is captured in the following lemma:
\begin{lemma}
\label{lem:LWE_dist}
Based on hardness assumptions of several lattice problems, the LWE distribution $A_{\mathbf{s},\bar{\Psi}_\alpha}$ of $(\mathbf{a},b)$'s is indistinguishable from a uniform distribution in $\mathbb{Z}_q^n\times\mathbb{Z}_q$.
\end{lemma}

Solving the decision version of LWE problem is to distinguish with a non-negligible advantage between samples from $A_{\mathbf{s},\bar{\Psi}_\alpha}$ and those generated uniformly from $\mathbb{Z}_q^n\times\mathbb{Z}_q$. 
This LWE problem is shown to be intractable to solve, without knowing the secret $\mathbf{s}$, based on the worst-case hardness of several lattice problems \cite{regev2009lattices}.
Errors $e$ are generated from a discrete Gaussian distribution $\bar{\Psi}_\alpha$ on $\mathbb{Z}_q$ parameterized by $\alpha >0$: sampling a continuous Gaussian random variable with mean $0$ and standard deviation $\alpha q/\sqrt{2\pi}$ and rounding it to the nearest integer in modulo $q$.  
Notice that error terms are also essential for guaranteeing the indistinguishability: without noise $(\mathbf{a},b)$ becomes deterministic and the secret $\mathbf{s}$ can be solved efficiently via Gaussian elimination methods.

We now describe a public-key cryptosystem based on the LWE problem above in \cite{brakerski2013classical}:
\begin{definition}{(LWE cryptosystem)}
\label{def:LWE_crypto}
\begin{itemize}
\item \textbf{Private key:} $\mathbf{s}$ is uniformly random in $\mathbb{Z}_q^n$ .
\item \textbf{Public key:}  $\mathbf{A}\in \mathbb{Z}_q^{m\times n}$ is uniformly random, and $\mathbf{e}\in \mathbb{Z}_q^m$ with each entry from $\bar{\Psi}_\alpha$. Public key is $(\mathbf{A},\mathbf{b}=\mathbf{A}\mathbf{s}+\mathbf{e})$.
\item \textbf{Encryption:}  $\mathbf{x}\in\{0,1\}^m$ is uniformly random. To encrypt a one-bit plaintext $r$, output ciphertext $\mathbf{c}=(\mathbf{a},b) = (\mathbf{A}^T\mathbf{x},\mathbf{b}^T\mathbf{x}+ r\floor{q/2})$.
\item \textbf{Decryption:} Decrypt the ciphertext $(\mathbf{a},b)$ to $0$ if $b-\innerprod{\mathbf{a},\mathbf{s}}$ is closer to $0$ than to $\floor{q/2}$ modulo $q$, and to $1$ otherwise. 
\end{itemize}
\end{definition}
Notice that each row in the public-key $(\mathbf{A},\mathbf{b})$ is an instance from the LWE distribution $A_{\mathbf{s},\bar{\Psi}_\alpha}$.

Correctness of the LWE cryptosystem can be easily verified: without the error terms, $b-\innerprod{\mathbf{a},\mathbf{s}}$ is either $0$ or $\floor{q/2}$, depending on the encrypted bit. 
Semantic security of the LWE cryptosystem follows directly from the indistinguishability of the LWE distribution from the uniform distribution in $\mathbb{Z}_q^n\times\mathbb{Z}_q$. 
Ciphertexts $(\mathbf{a},b)$ are either linear combinations or shifted linear combination of LWE samples, both of which are indistinguishable from the uniform distribution. 
This is true because shifting by any fixed length preserves the shape of a distribution.
Therefore, an efficient algorithm that can correctly guess the encrypted bit would be able to distinguish LWE samples from uniformly distributed samples. 
This allows \cite{regev2009lattices} to prove that:
\begin{theorem}
\label{thm:LWE_hardness}
Based on the hardness assumptions of several lattice problems, the LWE cryptosystem is secure against the chosen-plaintext attacks using both classical and quantum computers.
\end{theorem}

When the error terms $e_i$'s are introduced:
\begin{align*}
b-\innerprod{\mathbf{a},\mathbf{s}} = &\sum_{i\in S}b_i + \floor{\frac{q}{2}}r - \innerprod{\sum_{i\in S}\mathbf{a}_i,\mathbf{s}}\\
= &\sum_{i\in S}(\innerprod{\mathbf{a}_i,\mathbf{s}}+e_i) - \floor{\frac{q}{2}}r - \innerprod{\sum_{i\in S}\mathbf{a}_i,\mathbf{s}}\\
= &\floor{\frac{q}{2}}r - \sum_{i\in S}e_i,
\end{align*}
in which $S$ is the set of non-zero coordinates in $\mathbf{x}$.
For a decryption error to occur, the accumulated error $\sum_{i\in S}e_i$ must be greater than the decision threshold $\floor{q/4}$. 
The probability of the error is given by \cite{micciancio2009lattice}: 
\begin{align*}
\text{Err}_{\text{LWE}} &\approx 2(1-\Phi(\frac{q/4}{\alpha q \sqrt{m/2}/\sqrt{2\pi}})) \\
&= 2(1-\Phi(\frac{\sqrt{\pi}}{2\alpha\sqrt{m}})),
\end{align*}
in which $\Phi(\cdot)$ is the cumulative distribution function of the standard Gaussian variable. 
We later use this expression to find the practical parameters for the lattice PUF.

\section{Design of Lattice PUF}
\label{sec:design}
\begin{figure*}[t!]
    \centering
    \includegraphics[width = 0.68\linewidth]{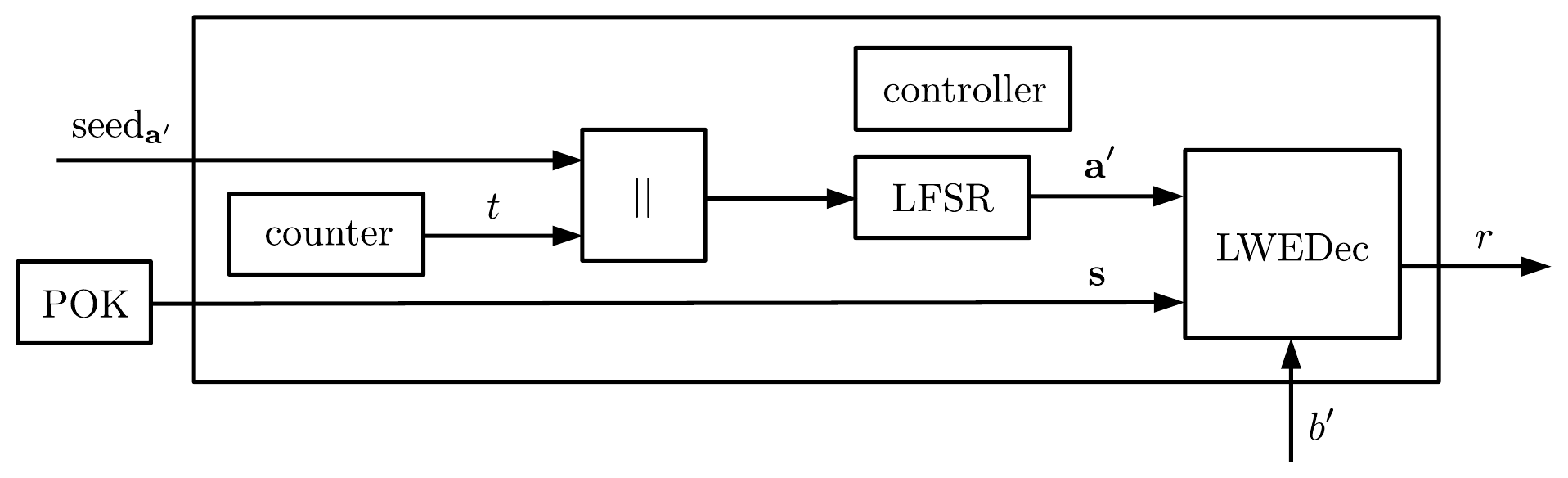}    \caption{Top-level architecture and data flow of the lattice PUF.}
    \label{fig:fpga_impl}
\end{figure*}
The theoretical security guarantees in Section \ref{sec:lwe} shows that an LWE decryption function can be used as a strong PUF with challenges generated from a ciphertext distribution. 
In this section, we first derive design parameters for the LWE decryption function and show that such a direct implementation of lattice PUF is inefficient in resource constrained environments due to high-ratio of ciphertext to plaintext. 
As we will illustrate in the following, an LWE decryption function with a 128-bit concrete ML hardness requires transmitting $128.8K$ challenge bits in order to produce a $100$-bit response string. 
We then solve this problem \emph{by exploiting distributional relaxations allowed by recent work in space-efficient LWEs}.
The proposed strategy allows introducing a low-cost PRNG based on an LFSR and transmitting only a small seed, which results in a dramatic reduction of effective challenge size. 
Last, we introduce a simple defense to protect our PUF against a standard active attack on the LWE decryption function. 

The top-level architecture of the proposed lattice PUF is shown in Figure \ref{fig:fpga_impl}.

\subsection{LWE Decryption Function}
\label{sec:lwe_dec}
\begin{figure}[t!]
    \centering
    \includegraphics[width = 0.9\linewidth]{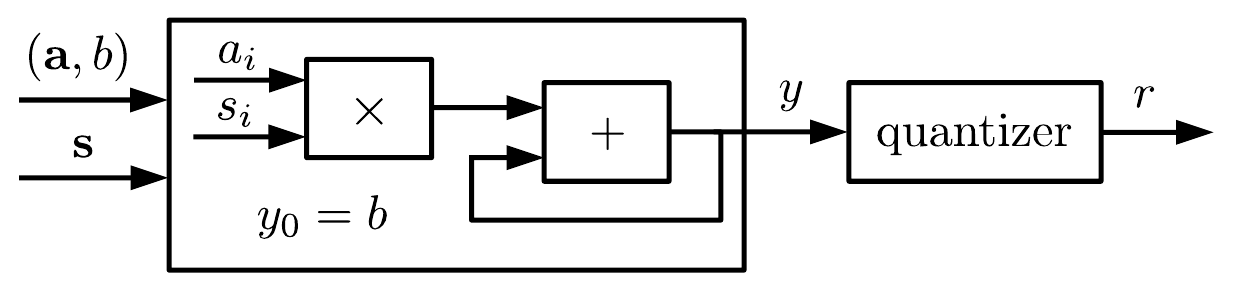}
    \caption{Architecture of LWE decryption function.}
    \label{fig:lwedec}
\end{figure}

Figure \ref{fig:lwedec} shows the core component of the proposed lattice PUF: the LWE decryption function. 
It takes a binary challenge vector $\mathbf{c} = \{c_0,c_1,\ldots,c_{N-1}\}$ of size $N = (n+1)\log q$ which maps to a ciphertext $(\mathbf{a},b)$ in the following way:
\begin{align*}
a_i &= \sum_{j=0}^{\log q-1}c_{(i-1)\log q+j}2^j,\; \forall i\in \{1,2,\ldots,n\}, \\
b &= \sum_{j=0}^{\log q-1}c_{n\log q+j}2^j. 
\end{align*}
Here $a_i$ denotes the $i$-th element of the integer vector $\mathbf{a}\in\mathbb{Z}_q^n$.
In this paper, without specification, $\log(x)$ refers to $\log_2(x)$.
Similarly, the private key $\mathbf{s}$ for the corresponding LWE decryption function is realized by a binary secret key $\mathbf{W} =\{W_0,W_1,\ldots,W_{n\log q-1}\} $ of size $n\log q$:
\begin{equation*}
s_i=\sum_{j=0}^{\log q-1} W_{(i-1)\log q+j} 2^j,\; \forall i\in \{1,2,\ldots,n\}.
\end{equation*}
A modulo-dot-product $b-\innerprod{\mathbf{a},\mathbf{s}}$ is computed using the modulo-multiply-accumulate unit. 
It can be implemented in a serial way using $n$ stages. 
Recall that all additions and multiplications are performed in modulo $q$.
Since $q$ is a power of $2$ in our construction, modulo addition and multiplication can be naturally implemented by integer addition and multiplication that keep only the last $\log q$-bit result. 
Finally the response $r$ is produced by a quantization operation $r = Q(b-\innerprod{\mathbf{a},\mathbf{s}})$: 
\begin{equation*}
Q(x) = \begin{cases}
	0& x \in [0,\frac{q}{4}]\cup(\frac{3q}{4},q-1],\\
	1& x \in (\frac{q}{4},\frac{3q}{4}].
\end{cases}
\end{equation*}

The computation above can be directly implemented as a strong PUF with $2^N$ CRPs since it maps a challenge vector $\mathbf{c}\in \{0,1\}^N$ into a binary response $r\in\{0,1\}$.
We now discuss parameter selection for the LWE decryption function. 
In general, we seek to find design parameters such that 
(1) the resulting PUF has excellent statistical properties, such as uniformity, uniqueness, and reliability, 
(2) successful ML attacks against it require an un-affordably high time complexity in practice, and 
(3) its hardware implementation costs are minimized. 

Prior theoretical arguments establish the impossibility of a polynomial-time attacker. 
To guarantee practical security, we need to estimate the number of samples and the actual running time (or a number of CPU operations) required for a successful ML attack. \cite{regev2009lattices} shows that a small number of samples are enough to solve an LWE problem, but in an exponential time. 
Thus, we refer to runtime as concrete ML resistance (or ML hardness) and say that a PUF has $k$-bit ML resistance if any successful ML attack requires at least $2^k$ operations. 
We adopt the estimator developed by Albrecht \emph{et al.} \cite{albrecht2015concrete} to estimate concrete ML hardness. 
The concrete hardness of an LWE problem increases with the increase of LWE parameters $n$, $q$, and $\alpha$ for all types of attacks.
Recall that $n$ represents the lattice dimension, $q$ represents the range of integer for each dimension, and $\alpha$ reflects the noise level in CRP (ciphertext) generation.
For a given set of parameters, the estimator compares the complexity of several most effective attacks, including decoding, basis reduction, and meet-in-the-middle attacks \cite{chen2011bkz,howgrave2007hybrid,lindner2011better}. 
We utilize the estimator in a black-box fashion to find the set of parameters with the target of $128$-bit concrete ML resistance. 

We consider two metrics of implementation cost, both of which scale with $n$: the number of challenge and secret bits needed ($n\log q$), and the number of multiply-accumulate (MAC) operations ($n$).
This motivates the need to decrease $n$.

For conventional PUFs, such as APUF and SRAM PUF, an output error is due to environmental noise, e.g. delay changes in APUF and FET strength changes in SRAM PUF with both voltage and temperature.
In contrast, output errors of the lattice PUF come from two sources: (1) environmental errors of secret bits, and (2) errors of decryption during response generation.
The former can be thought as the failure of key reconstruction in POKs.
Since a single bit-flip completely changes the challenge-response behavior of LWE decryption function, the failure rate of key reconstruction needs to be low, e.g. $10^{-6}$ (as widely adopted in other PUF applications \cite{maes2012pufky}).
Section \ref{sec:result} describes how the target failure rate can be achieved via a conventional FE based on the error-correcting codes.
The latter corresponds to the decryption error and is orthogonal to errors in the secret key $\mathbf{s}$. 
Recall that in CRP generation of the lattice PUF, a bit of plaintext $r$ is sampled and the ciphertext $\mathbf{c}$ is produced by a noisy encryption function $\mathbf{c}=\Enc(r)$. 
Given ciphertext $\mathbf{c}$ as input challenge, the decryption function can output a wrong response $r^\prime\neq r$ when the accumulated error $\sum_{i\in S} e_i$ in the encryption function exceeds the decision boundary.

\begin{figure}[t!]
\centering
\includegraphics[width = 0.93\linewidth]{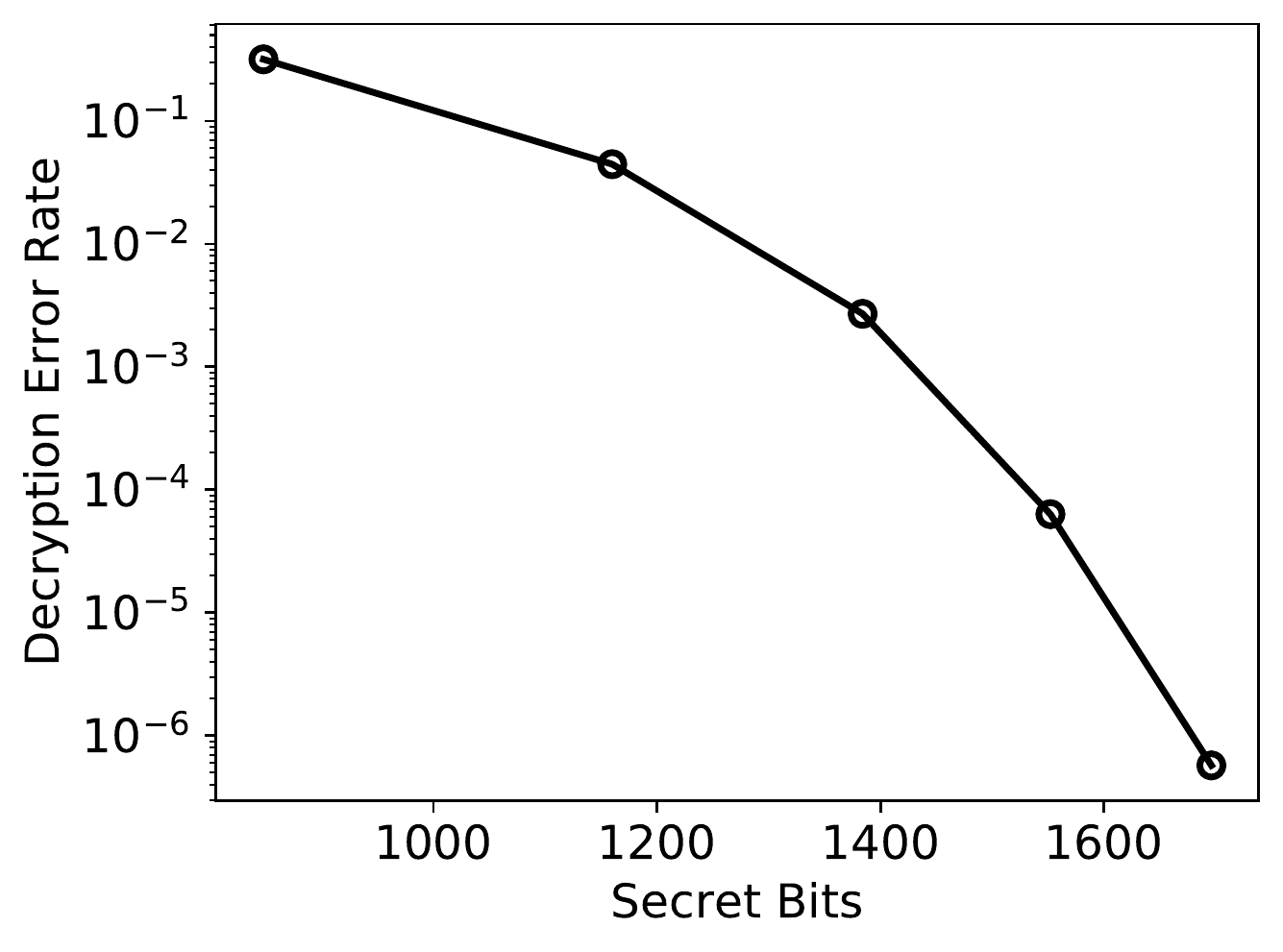}
\caption{Super-exponential decrease of decryption error rate with the increase of secret bits.
The analysis is done for 128-bit concrete hardness.}
\label{fig:num_POK_vs_dec_err}
\end{figure}

The model for evaluating the decryption error rate is shown in Section \ref{sec:bg}.
In order for a strong PUF to be used in direct authentication, its decryption error rate should be small enough for reliable distinguishability of long strings.
We set the target around $2\%$.
Figure \ref{fig:num_POK_vs_dec_err} explores the trade-off between the number of secret bits and the decryption error rate needed for $128$-bit concrete ML hardness.
It shows that, at fixed concrete ML hardness, the decryption error rate decreases super exponentially with the number of secret bits. 

Considering the design metrics above, a feasible set of parameters is found using the estimator in \cite{albrecht2015concrete}.
By setting $n=160$, $q=256$, $m=256$ and $\alpha = 2.20\%$, we achieve a lattice PUF with $128$-bit concrete hardness and a decryption error rate of $1.26\%$.

In order to get a $1$-bit response, $(n+1)\log q = 1288$ bits need to be sent to the lattice PUF as a challenge.
For direct authentication applications, usually around $100$ bits of responses are required. Therefore, the direct implementation described so far would require $C = 128.8K$ challenge bits.
This high ratio of challenge length to response length limits its practical use in many scenarios when communication is expensive. 

\subsection{Challenge Compression through Distributional Relaxation}
\label{sec:lfsr}
We now describe the proposed strategy based on space-efficient LWE that overcomes the limitation on communication inefficiency.
The LWE decryption function described in Section \ref{sec:lwe_dec} requires a challenge $\mathbf{c}$ in the form $\mathbf{c}=(\mathbf{a},b)$ to be sent from the server to the PUF. 
To represent vector $\mathbf{a} \in \mathbb{Z}_q^n$ requires $n\log q$ bits while to represent scalar $b \in \mathbb{Z}_q$  requires only $\log q$ bits. 
Thus, the major cost of transmission lies in sending $\mathbf{a}$. 
We wish to avoid sending $\mathbf{a}$ directly and, instead, to send a compressed (shorter) version of $\mathbf{a}$  and re-generate its full-size version on the PUF.
Our approach is enabled by the recent results on the distributional behavior of $\mathbf{a}=\mathbf{A}^T\mathbf{x}$ \cite{akavia2009simultaneous} and the concept of space-efficient LWE \cite{galbraith2013space}.

Recall that $b$ is given by: 
\begin{align*}
    b&=\mathbf{b}^T\mathbf{x}+r\floor{q/2}\\
    &= (\mathbf{A}\mathbf{s}+\mathbf{e})^T\mathbf{x}+r\floor{q/2}\\
    &=(\mathbf{A}^T\mathbf{x})^T\mathbf{s}+\mathbf{e}^T\mathbf{x}+r\floor{q/2}.
\end{align*}
First, we replace the component $\mathbf{a}=\mathbf{A}^T\mathbf{x}$ by $\mathbf{a}^*$ uniformly randomly sampled from $\mathbf{Z}_q^n$. That allows us to represent challenge $\mathbf{c} = (\mathbf{a},b)$:
\begin{equation*}
    \begin{cases}
    \mathbf{a}= \mathbf{A}^T\mathbf{x}\\
    b = (\mathbf{A}^T\mathbf{x})^T\mathbf{s}+\mathbf{e}^T\mathbf{x}+r\floor{q/2}
    \end{cases}
\end{equation*}
as $\mathbf{c}^*=(\mathbf{a}^*,b^*)$:
\begin{equation*}
    \begin{cases}
    \mathbf{a}^*\\
    b^*=\mathbf{a}^{*T}\mathbf{s}+\mathbf{e}^T\mathbf{x}+r\floor{q/2}
    \end{cases}.
\end{equation*}
In \cite{akavia2009simultaneous}, it is proven that distribution of $\mathbf{c}^*=(\mathbf{a}^*,b^*)$ is statistically close to the original ciphertext distribution, therefore the required security properties are preserved.  

The advantage of the above approximation is that, as shown by \cite{galbraith2013space}, several low-complexity  PRNGs are capable of producing an output string $\mathbf{a}^\prime$ suitably close to $\mathbf{a^*}\in\mathbb{Z}^n_q$ within the context of LWE cryptosystem. In particular, an LFSR is an especially simple PRNG having the right properties.
Specifically, a vector $\mathbf{a}^\prime$ generated by an LFSR provides similar concrete security guarantees against standard attacks on LWE, such as CVP reduction, decoding, and basis reduction \cite{galbraith2013space}.
This is because LFSR-generated $\mathbf{a}^\prime$ maintains good properties including:
\begin{itemize}
    \item it is hard to find ``nice'' bases for a lattice with basis from LFSR-generated $\mathbf{a}^\prime$;
    \item given an arbitrary vector in $\mathbb{Z}_q^n$, it is hard to represent it as a binary linear combination of LFSR-generated $\mathbf{a}^\prime$'s;
    \item it is hard to find a short vector $\mathbf{w}$ that is orthogonal to LFSR-generated $\mathbf{a}^\prime$'s.
\end{itemize}

The ability to rely on a simple PRNG to produce $\mathbf{a}^\prime$ allows a dramatic reduction in challenge transfer cost. 
Now, the challenge $\mathbf{c}^\prime$ contains only a small $\text{seed}_{\mathbf{a}^\prime}$ into the PRNG and the corresponding $b^\prime$ as
\begin{align*}
    b^\prime&=(\mathbf{a}^\prime)^T\mathbf{s}+\mathbf{e}^T\mathbf{x}+r\floor{q/2}\\
    &= \LFSR{(\text{seed}_{\mathbf{a}^\prime})}^T\mathbf{s}+\mathbf{e}^T\mathbf{x}+r\floor{q/2}.
\end{align*}
Here $\LFSR(\cdot)$ denotes the output generated by an LFSR.

With LWE parameters chosen as Section \ref{sec:lwe_dec}, using a seed of length $l=256$ is able to reduce the challenge length from $1288$ to $256+8=264$ per one bit of response. 
The improvement of efficiency becomes more pronounced for generating multiple responses:
This is because $\mathbf{a}^\prime_1 \ldots \mathbf{a}^\prime_t$ can be generated sequentially from the $l$-bit seed, so that only the seed and $b^\prime_1,\ldots, b^\prime_t \in Z_q$ are required to be sent to the PUF side. 
$100$ bits of responses now require only transmitting $256+100\times\log 256 = 1056$ bits for challenges.

\subsection{Countermeasure for Active Attack}
\label{sec:counter}
The focus of the paper is a PUF secure against passive attacks in which the observed challenges can be used to derive an internal model of the PUF. 
However, the LWE decryption function is vulnerable to an active attack that supplies arbitrary input challenges. 
(As we show, this risk also carries into an LFSR-based variant). 

The attack is premised on the ability to supply arbitrary challenges (ciphertexts) as inputs to the decryption function. 
The attack proceeds as follows. 
The attacker fixes $\mathbf{a}$ and enumerates all possible $b\in \mathbb{Z}_q$ for challenge $\mathbf{c} = (\mathbf{a},b)$.
As $b$ increases from $0$ to $q-1$, the response $r = Q(b-\innerprod{\mathbf{a},\mathbf{b}})$ changes from  $Q(b-\innerprod{\mathbf{a},\mathbf{s}}) = 0$ to $Q(b+1-\innerprod{\mathbf{a},\mathbf{s}}) = 1$ exactly when $b$ satisfies
\begin{equation*}
b-\innerprod{\mathbf{a},\mathbf{s}} = q/4.
\end{equation*}
We denote this specific value of $b$ as $\hat{b}$. 
The exact value of $\innerprod{\mathbf{a},\mathbf{s}}$ can then be extracted by $\innerprod{\mathbf{a},\mathbf{s}} = \hat{b} - q/4$. 
By repeating this procedure $n$ times, the attacker is able to set up $n$ linear equations (without errors):  
\begin{align*}
\label{eq:secret_eqs}
    \innerprod{\mathbf{a}_0,\mathbf{s}} &= \hat{b}_0 - q/4, \\
    \innerprod{\mathbf{a}_1,\mathbf{s}} &= \hat{b}_1 - q/4, \\
    & \cdots \\
    \innerprod{\mathbf{a}_{n-1},\mathbf{s}} &= \hat{b}_{n-1} - q/4.
\end{align*}
Gaussian elimination can then be used to solve for $\mathbf{s}$. 
The reason the attack succeeds is that attackers are able to fix $\mathbf{a}$ and use it for multiple values of $b$. 

We overcome the risk of such an attack by adopting the technique in \cite{yu2016lockdown}: we  introduce a self-incrementing counter to embed the counter value into a challenge seed. 
This makes the attack impossible as the counter restricts the attacker's ability to completely control input challenges to the LWE decryption function.
As a result, the attacker cannot enumerate all values of $b$ while keeping $\mathbf{a}$ unchanged. 
As shown in Figure \ref{fig:fpga_impl}, the concatenation of the challenger-provided seed and the counter value $t$ (i.e. $\text{seed}_{\mathbf{a}'}||t$) is used as the seed for generating $\mathbf{a}$. 
The counter value is public and is incremented by $1$ on each response generation.

\section{Experimental Results}
\label{sec:result}
In this section, we build a behavior model simulator of the constructed lattice PUF, in which the statistical model of raw SRAM POKs follows from \cite{maes2013accurate, maes2009soft} and other digital circuit components are accurately emulated by Python. 
$1000$ lattice PUF instances are virtually manufactured (simulated) and their CRPs are extracted to evaluate (1) statistical properties of the lattice PUF, including uniformity, uniqueness, and reliability with parameters chosen in Section \ref{sec:design}, and (2) vulnerability to state-of-the-art ML attacks. 
In order to quantize the lightweightness, we implement the entire lattice PUF system (except for the raw SRAM cells) on a Spartan 6 FPGA and compare it with prior work.

\subsection{Statistical Analysis}
\label{sec:statistical_result}
\begin{figure}[t!]
\centering
\includegraphics[width = 0.85\linewidth]{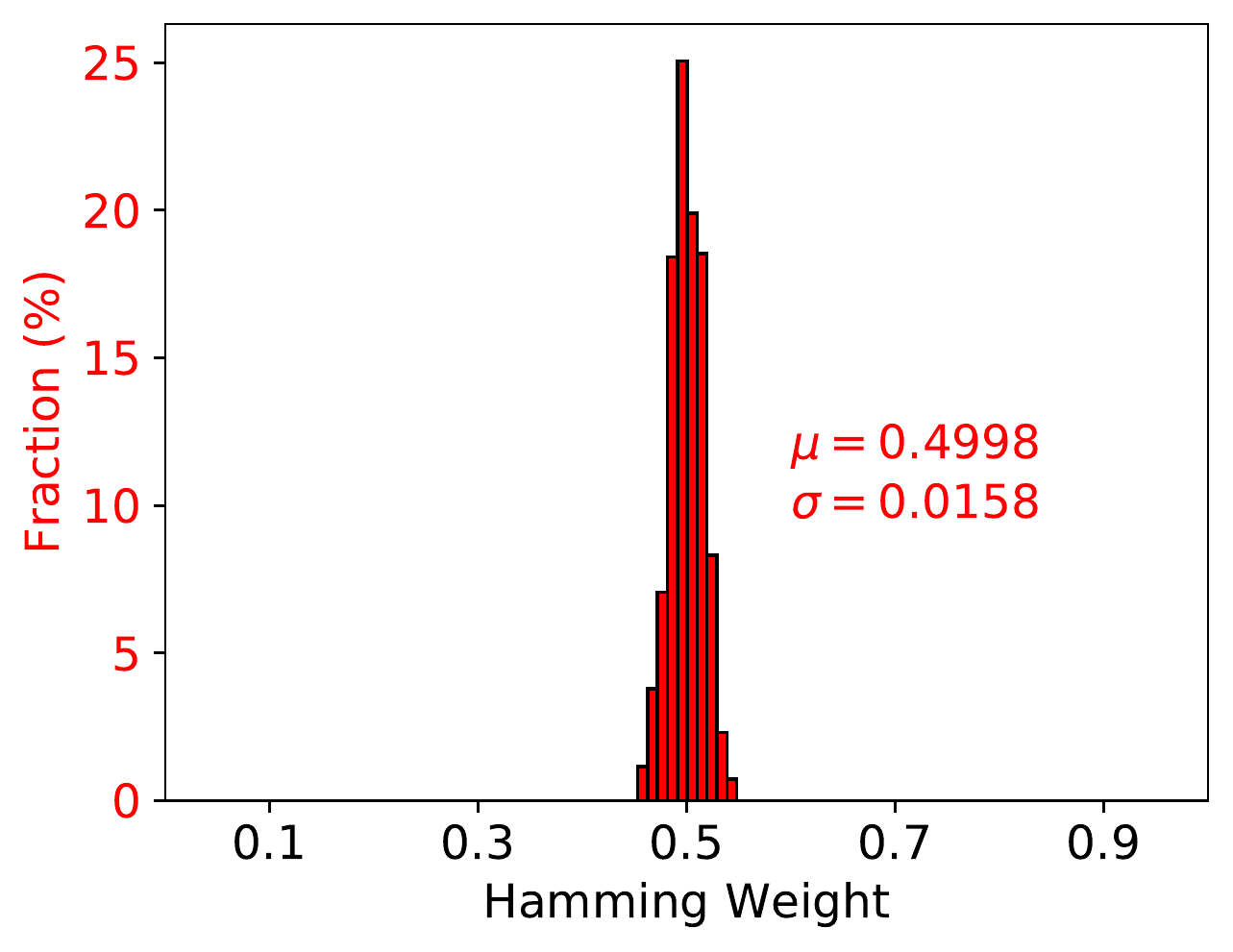}
\caption{Uniformity of lattice PUF output.}
\label{fig:HW}
\end{figure}

\textbf{Uniformity} of a PUF characterizes unbiasedness, namely, the proportion of `0's and `1's in the output responses.
For an ideal PUF $f$, the proportion needs to be $50\%$.
We adopt the definition of uniformity in \cite{maiti2013systematic} based on the average Hamming weight $\HW(f)$ of responses $\mathbf{r}$ to randomly sampled challenges $\mathbf{c}$'s:
\begin{equation*}
\HW(f)= \Expect_\mathbf{c}[\HW(\mathbf{r})] = \Expect_{\mathbf{c}}[\HW(f(\mathbf{c}))].
\end{equation*}
Here $\Expect_X$ represents expectation over random variable $X$.
Note that $\mathbf{c}$ follows the ciphertext distribution rather than the usual uniform distribution \cite{maiti2013systematic}.
Figure \ref{fig:HW} shows uniformity obtained using $1000$ randomly selected challenges. 
The distribution is centered at $49.98\%$, the standard deviation is $1.58\%$.

\textbf{Uniqueness} measures the ability of a PUF to be uniquely distinguished among a set of PUFs. 
Based on \cite{maiti2013systematic}, we define this metric to be the average inter-class HD of responses $(\mathbf{r}_i,\mathbf{r}_j)$ under the same challenges $\mathbf{c}$ for a randomly picked PUF pair $(f_i, f_j)$:
\begin{equation*}
\HD(f_i,f_j) = \Expect_\mathbf{c}[\HD(\mathbf{r}_i,\mathbf{r}_j)]=\Expect_{\mathbf{c}}[\HD(f_i(\mathbf{c}),f_j(\mathbf{c}))].
\end{equation*}
For ideal PUFs, responses under the same challenges are orthogonal, namely, $\HD(f_i,f_j)$'s are close to $50\%$.
Uniqueness is also evaluated under the ciphertext distribution.  

\begin{figure}[t!]
\centering
\includegraphics[width = 0.95\linewidth]{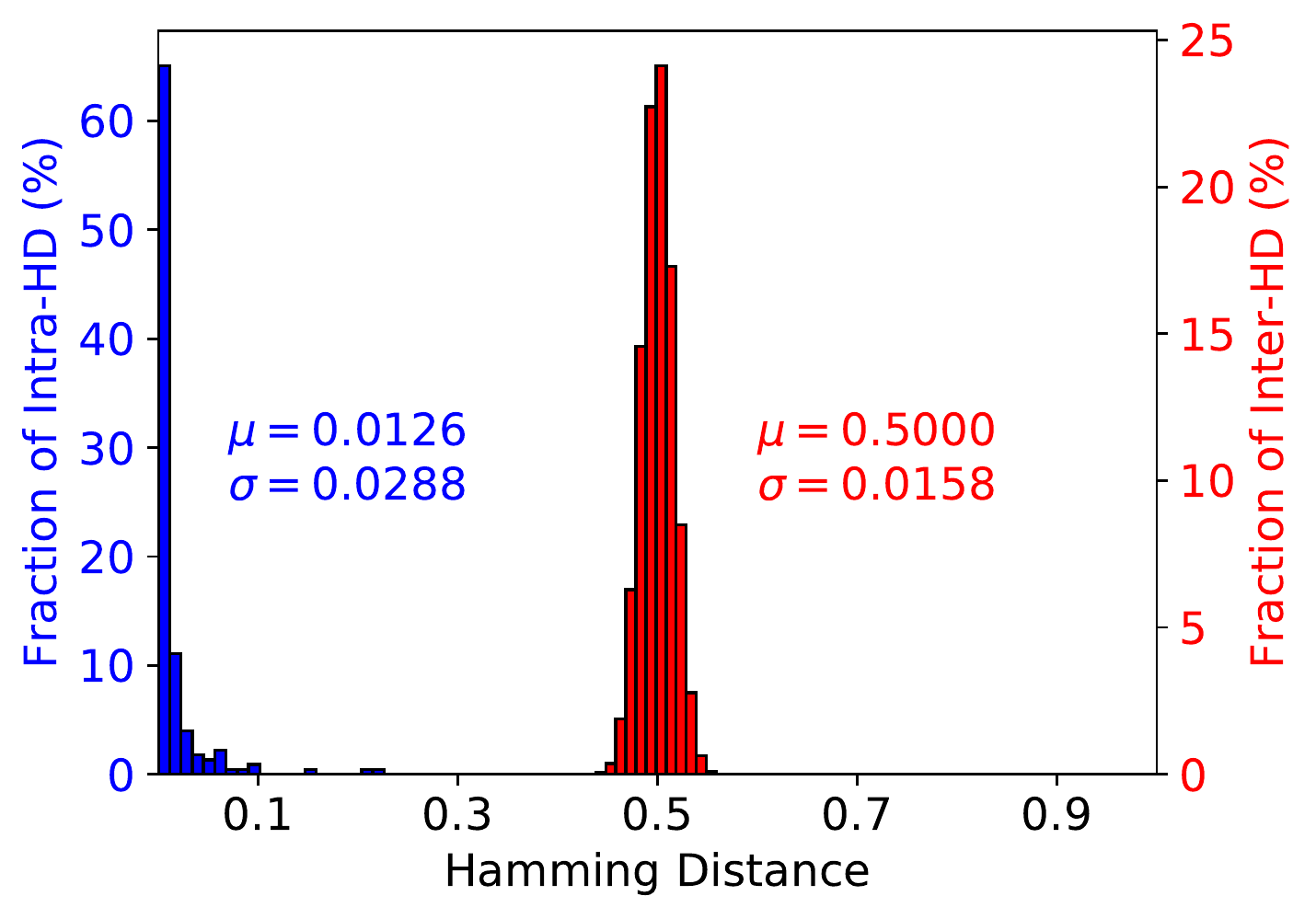}
\caption{Uniqueness and reliability of lattice PUF output.}
\label{fig:inter_intra}
\end{figure}

Uniqueness is shown in Figure \ref{fig:inter_intra}, evaluated for $1000$ PUF instances. 
The lattice PUF achieves near-optimal uniqueness: inter-class HD is centered at $50.00\%$, its standard deviation is $1.58\%$. 

\textbf{Reliability} of a PUF $f$ is characterized by the average BER of outputs with respect to their enrollment values:
\begin{equation*}
\BER = \Expect_{f^\prime}[\HD(f,f^\prime)]= \Expect_{f^\prime,\mathbf{c}}[\HD(f(\mathbf{c}),f^\prime(\mathbf{c}))].
\end{equation*}
As discussed in Section \ref{sec:design}, the overall BER of the lattice PUF is due to two components: the failure rate of key reconstruction and LWE decryption error rate.
Intra-class HD in Figure \ref{fig:inter_intra} reflects the result of decryption errors by assuming a perfect key reconstruction. 

\subsection{Empirical ML Resistance}

While the ultimate promise of lattice PUF is due to its theoretically-supported reliance on hard computational problems, testing its resilience to empirical ML attacks is important and provides additional confidence about the lattice PUF. 
We evaluate the vulnerability of lattice PUF to a range of traditional (i.e., not based on deep learning) ML attack methods, including SVM, LR, and single-layer NN (1-NN), as well as a number of DNNs.
We use the Python package scikit-learn \cite{scikit-learn} to implement SVM and LR.  
The SVM uses a nonlinear kernel with radial basis functions (RBF).
The 1-NN model uses only one hidden feed-forward layer composed of 100 neurons, with the rectified linear unit (ReLU) as the activation function. 
Training of 1-NNs and subsequent DNNs are implemented using Keras \cite{chollet2015keras} with TensorFlow \cite{abadi2016tensorflow} backend. 

\begin{figure}[t!]
\centering
\includegraphics[width = 0.92\linewidth]{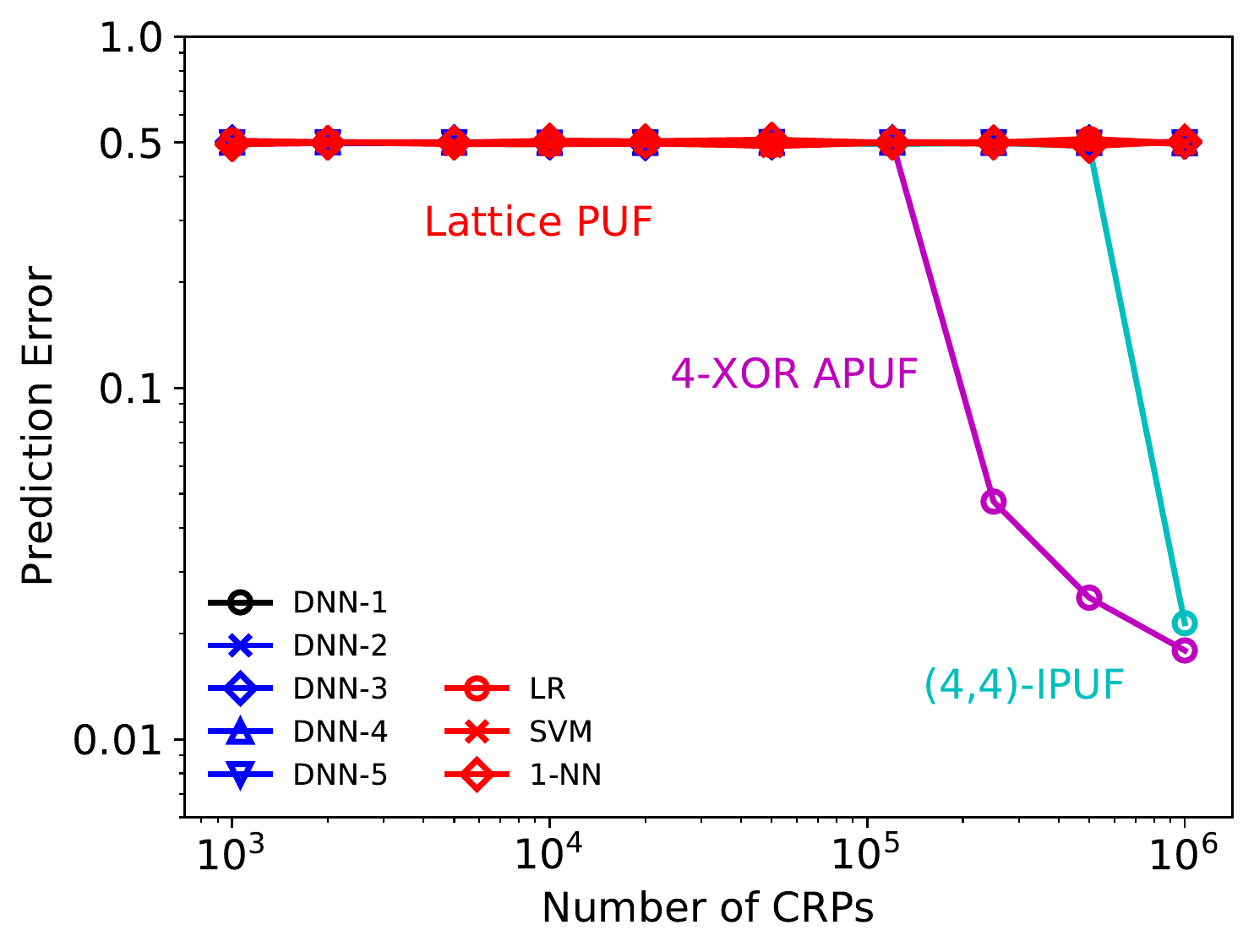}
\caption{ML attacks: Lattice PUF remains resistant to all attacks (DNNs, LR, SVM,1-NN). DNN ultimately succeeds in modeling two other strong PUFs.}
\label{fig:ml_attack_2}
\end{figure}

\begin{figure}[t!]
\centering
\includegraphics[width = 0.94\linewidth]{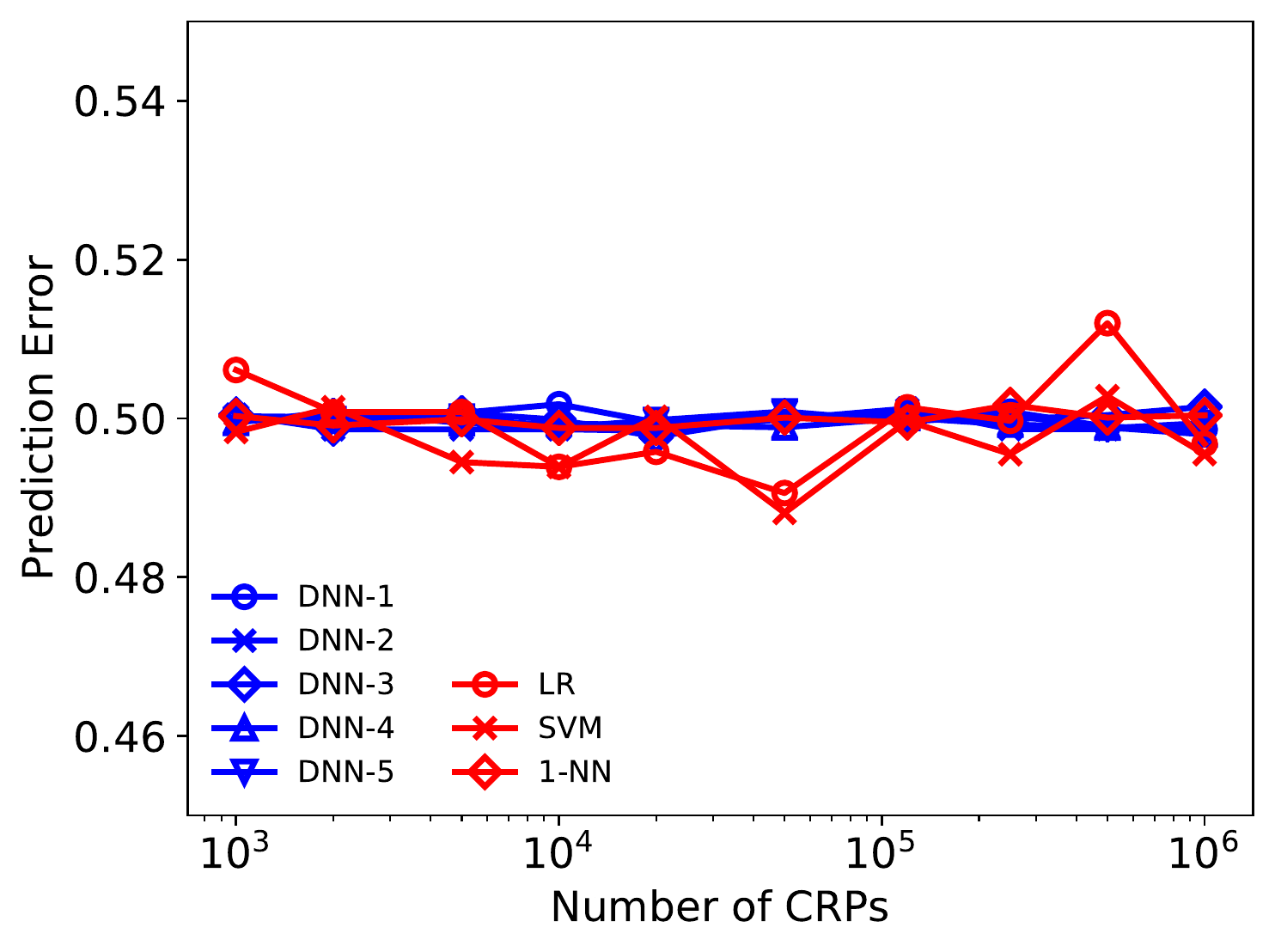}
\caption{Lattice PUF is resistant to both traditional ML attacks and DNNs.}
\label{fig:ml_attack_3}
\end{figure}

\begin{table}[t!]
\centering
	\caption{Various configuration for DNN attacks.}
	\label{table:DNNSetting}
\begin{tabular}{|l|l|l|l|l|l|}
\hline
Setup & \begin{tabular}[c]{@{}l@{}}Hidden\\ Layers\end{tabular} & \begin{tabular}[c]{@{}l@{}}Neurons\\ per Layer\end{tabular} & \begin{tabular}[c]{@{}l@{}}Challenge \\ Distribution\end{tabular} & \begin{tabular}[c]{@{}l@{}}Input \\ Format\end{tabular} & \begin{tabular}[c]{@{}l@{}}Prediction\\ Error\end{tabular} \\ \hline
DNN-1     & 4                                                       & 100                                                         & PRNG                                                              & Binary                                                  & 49.86\%                                                       \\ \hline
DNN-2     & 4                                                       & 100                                                         & PRNG                                                              & Real                                                    & 49.84\%                                                       \\ \hline
DNN-3     & 4                                                       & 100                                                         & Ciphertext                                                        & Binary                                                  & 49.76\%                                                       \\ \hline
DNN-4     & 6                                                       & 100                                                         & PRNG                                                              & Binary                                                  & 49.80\%                                                       \\ \hline
DNN-5     & 4                                                       & 200                                                         & PRNG                                                              & Binary                                                  & 49.87\%                                                       \\ \hline
\end{tabular}
\end{table}

DNNs represent a more powerful class of binary classifiers. DNNs contain multiple hidden layers that produce superior modeling expressiveness compared to 1-NNs \cite{goodfellow2016deep}. Importantly, DNNs have been recently shown to be very effective in attacking known strong PUFs, including XOR APUFs and IPUFs \cite{DBLP:journals/iacr/SantikellurBC19}. Our baseline DNN experiment (DNN-1) is based on the same network parameters as in \cite{DBLP:journals/iacr/SantikellurBC19}.  The network has 4 hidden layers, each containing 100 neurons. It uses ReLU as the non-linear operator. 

In addition to the baseline configuration, we explored attacks using several other network architectures, hyper-parameters, and numeric representations, as listed in Table \ref{table:DNNSetting}. 
DNN-2 treats input as 161 integer numbers from 0 to 255, instead of 1288 binary bits.
DNN-3 is different from the baseline version in its CRP generation strategy (see more below).
DNN-4 and DNN-5 add more hidden layers and more neurons per hidden layer respectively, compared to the baseline DNN-1.

Figure \ref{fig:ml_attack_2} shows the results of the empirical attacks based on the above ML algorithms. The figure shows the prediction error of lattice PUF in response to these attacks with training set size ranging from $1000$ to $1$ million and test set of size $200$K. The Adam optimizer \cite{kingma2014adam} terminates after 200 epochs, and results in a prediction error of $49.86\%$ for the proposed lattice PUF, barely better than a random guess. \emph{The results show that the prediction error of lattice PUF remains flat for all attempted attacks: across the range of attacks and CRP training set sizes, there is no measurable deviation of the error from 0.5.} In contrast, a DNN (with a configuration corresponding to DNN-1) achieves less than $2\%$ prediction error for both 4-XOR APUF and (4, 4)-IPUF. 

It is critical that the experiments also demonstrate that lattice PUF design that utilizes the distributional relaxation of space-efficient LWE (described in Section \ref{sec:lfsr}) shows the same empirical resistance to ML attacks as the design not based on such a relaxation. In Table 1, all design/attack combinations except for DNN-3 are based on the compact (relaxation-based) design in which CRPs are generated via a PRNG. Finally, we also provide an expanded view of the results of ML attacks on lattice PUF in Figure \ref{fig:ml_attack_3} by zooming in Figure \ref{fig:ml_attack_2}. While run-to-run variations of the training optimizer are observable, the prediction error remains close to $50\%$.

\subsection{Hardware Implementation Results}
\label{sec:hardware_results}

\begin{table}[t!]
    \centering
	\caption{Configuration of error-correcting codes.}
	\vspace{0.5em}
	\label{table:ecc}
	\def\arraystretch{1.1}
	\resizebox{0.9\linewidth}{!}{
        \begin{tabular}{|c|c|c|c|c|c|}
        \hline
        \multirow{2}{*}{\begin{tabular}[c]{@{}c@{}}\textbf{Raw BER}\\  \textbf{(\%)}\end{tabular}} & \multicolumn{2}{c|}{\textbf{Error-Correcting Code}}                                 & \multirow{2}{*}{\textbf{Raw POKs}}  \\ \cline{2-3} 
                                                                                 & Outer code   & Inner code & \\ \hline
        1                                                                        & {[}236, 128, 14{]}  & N/A            & 2,360  \\ \hline
        5                                                                        & {[}218, 128, 11{]} & {[}3, 1, 1{]}  & 6,540  \\ \hline
        10                                                                       & {[}220, 128, 12{]} & {[}5, 1, 2{]}  & 11,000  \\ \hline
        15                                                                       & {[}244, 128, 15{]} & {[}7, 1, 3{]}  & 17,080   \\ \hline
        \end{tabular}
    }
\end{table}

\begin{table}[t!]
    \caption{(a) Area consumption and (b) runtime of our reference lattice PUF implementation on Xilinx Spartan-6 FPGA.}
    \label{table:fpga_result}
    \centering
    \def\arraystretch{1.1}
    \subfloat[]{
        \resizebox{0.45\linewidth}{!}{
            \begin{tabular}{|c|c|}
            \hline
            \textbf{Module}         & \textbf{Size [slices]} \\ \hline
            LFSR                    & 27            \\ \hline
            LWEDec                  & 2             \\ \hline
            Controller              & 16            \\ \hline
            \textit{Total}          & 45            \\ \hline
            \end{tabular}
            \label{table:fpga_utilization}
        }
    }
    \hspace{1em}
    \subfloat[]{
        \resizebox{0.75\linewidth}{!}{
            \begin{tabular}{|c|c|}
            \hline
            \textbf{Step}                               & \textbf{Time [$\mu$s]} \\ \hline
            Seed $\text{seed}_{\mathbf{a}'}||t$ load for LFSR    & 8             \\ \hline
            1-bit decryption from LWEDec                  & 44            \\ \hline
            \textit{Total} @ 33 \textit{MHz}            & 52            \\ \hline
            \end{tabular}
            \label{table:fpga_timing}
        }
    }
\end{table}

\begin{table}[t!]
\centering
	\caption{Hardware implementation costs of strong PUFs.}
	\label{table:hardware_puf}
	\def\arraystretch{1.1}
	\resizebox{\linewidth}{!}{
        \begin{tabular}{|c|c|c|}
        \hline
        \textbf{Design}                                         & \textbf{Platform}  & \textbf{PUF Logic [Slices]} \\ \hline
        POK+AES \cite{bhargava2014efficient}                    & Spartan 6 & 80                 \\ \hline
        Controlled PUF \cite{gassend2008controlled}             & Spartan 6 & 127                \\ \hline
        CFE-based PUF \cite{herder2017trapdoor,jin2017fpga}  & Zynq-7000 & 9,825              \\ \hline
        Lattice PUF                                             & Spartan 6 & 45                 \\ \hline
        \end{tabular}
    }
    \vspace{-1em}
\end{table}

\begin{table*}[t!]
    \centering
    \caption{Hardware utilization in FE design on Spartan 6 FPGA.}
    \label{table:hardware_fe}
        \begin{tabular}{|c|c|c|c|c|c|c|c|c|c|}
        \hline
        \multirow{2}{*}{\begin{tabular}[c]{@{}c@{}}\textbf{Raw BER}\\  \textbf{(\%)}\end{tabular}} & \multicolumn{3}{c|}{\textbf{Outer Code}} & \multicolumn{3}{c|}{\textbf{Inner Code}} & \multicolumn{3}{c|}{\textbf{Total}} \\ \cline{2-10} 
                               & Reg      & LUT      & Slice     & Reg      & LUT      & Slice     & Reg    & LUT    & Slice    \\ \hline
        1                      & 905      & 893      & 276       & 0        & 0        & 0         & 905    & 893    & 276      \\ \hline
        5                      & 730      & 688      & 232       & 0        & 1        & 1         & 730    & 689    & 233      \\ \hline
        10                     & 785      & 740      & 243       & 0        & 3        & 2         & 785    & 743    & 245      \\ \hline
        15                     & 973      & 913      & 326       & 0        & 7        & 3         & 973    & 920    & 329      \\ \hline
        \end{tabular}
\end{table*}

\begin{figure}[t!]
\centering
\includegraphics[width = 0.75\linewidth]{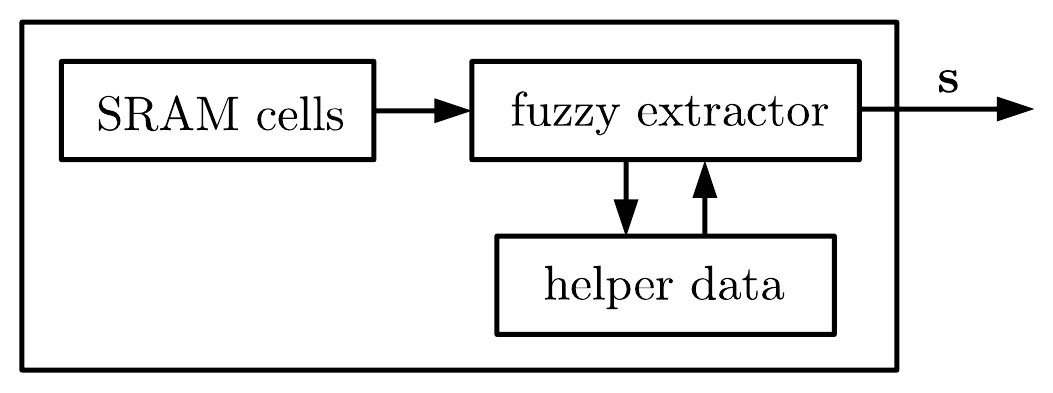}
\caption{POK uses an FE to ensure stability of the secret seed.}
\label{fig:pok}
\end{figure}

We now present the details of lattice PUF implementation and analyze its hardware efficiency.
The entire design, except for the raw (uninitialized) SRAM cells, was synthesized, configured, and tested on a Xilinx Spartan-6 FPGA (XC6SLX45), a low-end FPGA in 45nm technology.

Regarding the FE design, we adopt the homogeneous error assumption, i.e., all cells have the same BER \cite{bosch2008efficient}.
Prior work shows that intrinsic BERs of the various POKs range from $0.1\%$ \cite{karpinskyy20168} to $15\%$ \cite{maes2009soft}. 
We study the POK and FE designs under four levels of raw BER: $1\%$, $5\%$, $10\%$, and $15\%$ to explore design costs, and choose $5\%$ as the raw SRAM BER to benchmark our final implementation. 
Here the goal of FE design is to ensure a key reconstruction of $1280$ bits with targeted failure rate at $10^{-6}$.
As mentioned in Section \ref{sec:design}, with such nearly-perfect secret key reconstruction, the overall output BER of the lattice PUF can reach the decryption error rate analyzed above.
We use concatenated error-correcting codes, with a repetition code as the inner code, and a shortened BCH code as the outer code.
Concatenated codes are typically more efficient than single codes in terms of code length and hardware cost \cite{bosch2008efficient}.
Block diagram of the FE design is shown in Figure \ref{fig:pok}.
Table \ref{table:ecc} and \ref{table:hardware_fe} list the configuration and hardware costs of error-correcting codes used at different BER levels respectively.
At the raw BER of $5\%$, $6.5K$ cells are needed to construct the secret $\mathbf{s}$ of length $1280$ bits at the target failure rate $10^{-6}$.
The FE design of the lattice PUF requires 233 slices. 
This portion of cost applies to all other strong PUF candidates (AES PUF or other controlled PUF).
This is also cheaper than linear solver block used in the CFE-based strong PUF \cite{herder2017trapdoor,jin2017fpga} for key reconstruction, which requires $65,700$ LUTs and $16,425$ slices.

Regarding the PUF logic, the total size of the lattice PUF (without FE) for the Spartan-6 platform is 45 slices, most of which is taken up by the LFSR and the controller. 
Table \ref{table:fpga_utilization} shows the breakdown of resources needed to realize the various modules. 
The core block implementing the LWE decryption function (LWEDec) includes an 8-bit MAC and a quantization block, as shown in Figure \ref{fig:lwedec}. 
The 256-bit LFSR is implemented using RAM-based shift registers.
The total latency (at 33.3MHz clock) to generate a 1-bit PUF response is $47\mu$s, and the total time to generate a 100-bit PUF response is, approximately, $8\mu s + 100\times 44\mu s\approx 4.4ms$ since seed loading is only executed once. 
Table \ref{table:fpga_timing} lists the latency of each step of response generation. 

We compare the implementation cost of the lattice PUF logic against established strong PUF designs \cite{bhargava2014efficient, gassend2008controlled, jin2017fpga} in Table \ref{table:hardware_puf}. 
The original strong PUF based on AES \cite{bhargava2014efficient} is implemented as an ASIC. 
Here, we adopt \cite{chu2012low} as an FPGA alternative to estimate the implementation cost of AES. 
Notice that \cite{bhargava2014efficient} uses no error correction since it guarantees reliability via dark bit masking. 
Similarly, the FPGA implementation of SHA-3 \cite{kaps2011lightweight} is adopted to estimate the cost of a hash function for the controlled PUF \cite{gassend2008controlled}. 
The FPGA utilization result of the strong PUF based on the computational FE (CFE) is presented via the number of LUTs in \cite{jin2017fpga}. 
We estimate the corresponding slice count using \cite{xilinx:ds190}. 
Compared to PUFs based on AES \cite{chu2012low} and SHA \cite{kaps2011lightweight}, our advantages in area are minor. 
However, compared to \cite{herder2017trapdoor, jin2017fpga}, which is another PUF based on LWE, and which therefore provides similar theoretical guarantees, our savings in area are significant.

\section{Conclusion}
\label{sec:conclusion}
In this paper, we described a new strong physical unclonable function (PUF) that is provably secure against machine learning (ML) attacks with both classical and quantum computers. 
The security is derived from cryptographic hardness of learning decryption functions of semantically secure public-key cryptosystems within the probably approximately correct  framework.
The proposed PUF compactly realizes the decryption function of the learning-with-errors (LWE) public-key cryptosystem as the core block. 
We implemented a lattice PUF on a Spartan 6 FPGA. The design realizes a challenge-response pair space of size $2^{136}$, requires $1280$ physically obfuscated key bits, and guarantees $128$-bit ML resistance. 
The PUF shows excellent uniformity, uniqueness, and reliability. 

\section*{Acknowledgments}
We thank Dr.~Aydin Aysu for his insightful advice on idea presentation, assistance with FPGA implementation of repetition code, and comments that greatly improved the manuscript.

\bibliographystyle{abbrv}
\bibliography{refs/refs.bib}

\end{document}